\documentclass[twocolumn,showpacs,floatfix,prl]{revtex4}

\usepackage[dvips]{epsfig}

\usepackage{float}

\begin{document}

\title{Superconducting proximity effect and Majorana fermions
at the surface of a topological insulator
}

\author{Liang Fu and C.L. Kane}
\affiliation{Dept. of Physics and Astronomy, University of Pennsylvania,
Philadelphia, PA 19104}

\begin{abstract}
We study the proximity effect between an $s$-wave superconductor and the
surface states of a strong topological insulator.  The resulting two
dimensional state resembles a spinless $p_x+ip_y$ superconductor, but
does not break time reversal symmetry.  This state supports Majorana
bound states at vortices.  We show that linear junctions between
superconductors mediated by the topological insulator form a non chiral 1
dimensional wire for Majorana fermions, and that circuits formed from
these junctions provide a method for creating, manipulating and
fusing Majorana bound states.
\end{abstract}

\pacs{71.10.Pm, 74.45.+c, 03.67.Lx, 74.90.+n}
\maketitle

Excitations with non-Abelian statistics\cite{mooreread} are the basis for the
intriguing proposal of topological quantum computation\cite{kitaev}.  The simplest
non-Abelian excitation is the zero energy Majorana bound state (MBS)
associated with a vortex in a spinless $p_x+ip_y$
superconductor\cite{readgreen,ivanov,stern,stone}.  The presence of $2N$ vortices
leads to a $2^N$ fold
ground state degeneracy.  Braiding processes, in which the
vortices are adiabatically rearranged, perform non trivial operations
in that degenerate space.  Though MBSs do not have the structure necessary
to construct
a universal quantum computer\cite{friedman}, the quantum information encoded in
their degenerate states is topologically protected from local sources of
decoherence\cite{kitaev3}.

MBSs have been proposed to exist as quasiparticle
excitations of the $\nu=5/2$ quantum Hall effect\cite{mooreread,readgreen},
 in the cores of
$h/4e$ vortices in the $p$-wave superconductor Sr$_2$RuO$_4$\cite{dassarma}
and in cold atoms\cite{gurarie,tewari}.  In this
paper we show that the proximity effect between an ordinary $s$-wave
superconductor and the surface of a strong topological
insulator (TI)\cite{fkm,moore,roy,fukane}
leads to a state which hosts MBSs at vortices.  We then
show that a linear superconductor - TI - superconductor
(STIS) junction forms a non chiral 1D wire for Majorana fermions.
Such junctions can be combined into circuits, which allow for
the creation, manipulation and fusion of MBSs.

A strong TI is a material with an insulating
time reversal invariant
bandstructure for which strong spin orbit interactions lead to an
inversion of the band gap at an odd number of time reversed pairs of
points in the Brillouin zone.
Candidate materials include the semiconducting alloy
Bi$_{1-x}$Sb$_x$, as well as HgTe and $\alpha$-Sn under uniaxial
strain\cite{fukane}.  Strong TIs are distinguished from ordinary insulators
by the presence of surface states, whose Fermi arc encloses an odd number of
Dirac points and is associated with a Berry's phase of $\pi$.  In the
simplest case, there is a single non degenerate Fermi arc
described by the time reversal invariant Hamiltonian
\begin{equation}
H_0 = \psi^\dagger (-i v\vec\sigma\cdot \nabla -\mu)\psi.
\label{h0}
\end{equation}
Here $\psi = (\psi_\uparrow, \psi_\downarrow)^T$ are electron field
operators, $\vec \sigma =
(\sigma^x,\sigma^y)$ are Pauli spin matrices and $\mu$ is the
chemical potential.
$H_0$ can only exist on a surface because it violates the fermion
doubling theorem\cite{nielson}.  The topological metal is essentially {\it half}
of an ordinary 2D electron gas.

Suppose that an s-wave superconductor is deposited on the surface.
Due to the proximity effect, Cooper pairs can tunnel into the surface
states.  This can be described by adding
$V = \Delta \psi_\uparrow^\dagger \psi_\downarrow^\dagger + h.c.$ to
$H_0$, where $\Delta = \Delta_0 e^{i\phi}$ depends on the phase
$\phi$ of the superconductor and the nature of the interface\cite{volkov}.
The states of the surface can then be described by
$H = \Psi^\dagger {\cal H} \Psi/2$, where in the Nambu
notation $\Psi = ((\psi_\uparrow,
\psi_\downarrow),(\psi_\downarrow^\dagger,-\psi_\uparrow^\dagger))^T$ and
\begin{equation}
{\cal H} = - i v \tau^z \sigma\cdot\nabla - \mu \tau^z + \Delta_0
(\tau^x \cos\phi + \tau^y\sin\phi).
\label{hbdg}
\end{equation}
$\vec\tau$ are Pauli matrices that mix the $\psi$ and $\psi^\dagger$ blocks of $\Psi$.
Time reversal invariance follows from $[\Theta,{\cal
H}]=0$, where $\Theta = i \sigma^y K$
and $K$ is complex conjugation.  Particle hole symmetry
is expressed by $\Xi = \sigma^y \tau^y K$, which satisfies $\{\Xi,{\cal H}\}=0$.
When $\Delta$ is spatially homogeneous, the excitation spectrum is
$E_{\bf k} = \pm \sqrt{(\pm v|{\bf k}| - \mu)^2 + \Delta_0^2}$.  For
$\mu \gg \Delta_0$, the low energy
spectrum resembles that of a spinless $p_x+i p_y$ superconductor.
This analogy can be made precise by defining $c_{\bf k} =
(\psi_{\uparrow {\bf k}} + e^{i\theta_{\bf k}} \psi_{\downarrow {\bf
k}})/\sqrt{2}$ for ${\bf k} = k_0(\cos \theta_{\bf k},\sin\theta_{\bf
k})$ and $v k_0 \sim \mu$.  The projected Hamiltonian is then
$\sum_{\bf k} (v|{\bf k}|-\mu)  c_{\bf k}^\dagger c_{\bf k} +
(\Delta e^{i\theta_{\bf k}} c_{\bf k}^\dagger c_{-{\bf k}}^\dagger +
h.c.)/2$.  Though this is formally equivalant to a
spinless $p_x + i p_y$ superconductor there is an important
difference: ${\cal
H}$ respects time reversal symmetry, while the $p_x+i p_y$
superconductor does not.

It is well known that a $h/2e$ vortex in a $p_x+i p_y$ superconductor
leads to a MBS\cite{readgreen}.  This suggests that for
$\mu \gg \Delta_0$ a
similar bound state
should exist for (\ref{hbdg}).  The bound states at a
vortex are determined by solving the Bogoliubov de Gennes (BdG) equation
${\cal H} \xi = E \xi$ in polar coordinates
with $\Delta(r,\theta) =
\Delta_0(r) e^{\pm i\theta}$.  A zero energy solution exists
for any $\mu$.  The algebra is simplest for $\mu = 0$, where the zero
mode has the form
\begin{equation}
\xi^\pm_0(r,\theta) = \chi^\pm e^{-\int_0^r dr' \Delta_0(r')/v},
\label{vortex}
\end{equation}
with $\chi^+ = ((0,i),(1,0))^T$ and $\chi^-=((1,0),(0,-i))^T$.

Another feature of $p_x + i p_y$ superconductors is the
presence of chiral edge states
\cite{readgreen,buchholtz,sigrist}.
With time reversal symmetry,
chiral edge states can {\it not} occur in our system.  The
surface - which itself is the boundary of a three dimensional crystal
- can not have a boundary.  By breaking time reversal symmetry,
however, a Zeeman field can introduce a mass term $M\sigma^z$ into
(\ref{h0},\ref{hbdg})
which can open an insulating
gap in the surface state spectrum.  By solving (\ref{hbdg})
we find that the interface between this
insulating state and the superconducting state has chiral
Majorana edge states.  This could possibly be
realized by depositing superconducting and insulating magnetic materials
on the surface to form a superconductor-TI-magnet (STIM) junction.  It is interesting to note that
for spinless electrons the $p_x+ip_y$ superconductor violates
time reversal, while the vacuum does not.  For our surface states it is
the {\it insulator} which violates time reversal.
A related effect could also occur at the
edge of a {\it two dimensional} TI\cite{km,murakami,bhz},
which is described by (\ref{h0},\ref{hbdg}) restricted to one spatial dimension.
At the boundary between a region with superconducting gap $\Delta \tau^x$
and a region with insulating gap $M \sigma^z$ we find a
MBS, analogous to the end states
discussed in Refs. \onlinecite{kitaev2,semenoff}.
In the following we will focus on STIS junctions, which
can lead to {\it non chiral} one dimensional Majorana fermions, as well
as MBSs.

\begin{figure}
 \centerline{ \epsfig{figure=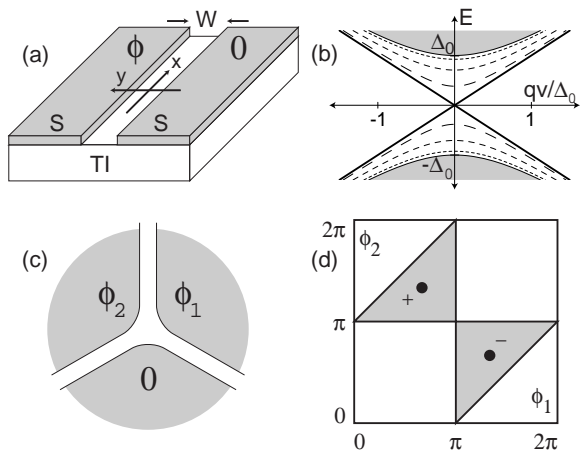,width=3.0in} }
 \caption{(a) A STIS line junction.  (b) Spectrum of a line junction
 for $W=\mu=0$ as a function of
 momentum for various $\phi$.  The solid line shows the Andreev bound
 states for $\phi=\pi$.  The dashed lines are for $\phi = 3\pi/4$,
 $\pi/2$ and $\pi/4$.  The bound states for $\phi=0$ merge with the
 continuum, indicated by the shaded region.
   (c) A tri-junction between three superconductors.  (d)
 Phase diagram for the tri-junction.  In
 the shaded regions there is a  $\pm$ MBS at the junction.
 }
\end{figure}

Consider a line junction of width $W$
and length $L\rightarrow \infty$
between two superconductors with phases $0$ and $\phi$ in contact with
TI surface states.
We analyze the Andreev bound states in the surface state channel
between the superconductors by solving the BdG
equation with $\Delta(x,y) = \Delta_0 e^{i\phi}$ for
$y>W/2$, $\Delta_0$ for $y<-W/2$ and $0$ otherwise.
The calculation is similar to Titov, Ossipov and
Beenakker's\cite{beenakker} analysis of graphene SNS junctions,
except for the
important difference that graphene has four independent Dirac points,
while we only have one.
For $W \ll v/\Delta_0$ there are two branches
of bound states, which disperse with the momentum
$q$ in the $x$ direction.  For  $W=\mu=0$
we find
\begin{equation}
E_\pm(q) = \pm \left[v^2 q^2 + \Delta_0^2 \cos^2 (\phi/2)\right]^{1/2}.
\end{equation}

For $\phi=\pi$ the spectrum is gapless. It is useful to construct a
low energy theory, for $q \sim 0$ and $\phi = \pi - \epsilon$. Finite
$W$ and $\mu$ can then easily be included. We first solve the BdG
equation for the two $E=0$ modes $\zeta_{a=1,2}(y)$ at $q=0$
and $\phi=\pi$.  It is useful to choose them to satisfy $\Xi \zeta_a
= \zeta_a$. Up to a normalization they may be written
\begin{equation} \zeta_1 \pm i
\zeta_2 =  ((1,\pm i),(\pm i,-1))^T
 e^{\pm i \mu y/ v -\int_0^{|y|}d\tilde y\Delta_0(\tilde y)/v}.
 \label{zeta}
\end{equation}
We next evaluate $\langle\zeta_a| q \sigma^x\tau^z|\zeta_b\rangle$ and
$\langle\zeta_a|\epsilon\Delta_0 \theta(y-W) \tau^y|\zeta_b\rangle$
to obtain the ``$k\cdot p$" Hamiltonian,
\begin{equation}
\tilde{\cal  H} = -i \tilde v \tau^x \partial_x + \delta \tau^y,
\label{h1d}
\end{equation}
where
$\tilde v = v[\cos\mu W +
(\Delta_0/\mu)\sin\mu W]\Delta_0^2/(\mu^2+\Delta_0^2)$ and
$\delta = \Delta_0 \cos(\phi/2)$.
The Pauli matrices $\tau^{x,y}_{ab}$ act on $\zeta_a$ and are different from
those in (\ref{hbdg}).  In this basis $\Theta = i\tau^y K$ and $\Xi = K$.
$\tilde{\cal H}$ resembles the Su Schrieffer Heeger (SSH)
model\cite{ssh}.
However, unlike that model, the $E_\pm(q)$
states are not independent, and
the corresponding Bogoliubov quasiparticle operators
satisfy $\gamma_+(q) = \gamma_-(-q)^\dagger$.
The system is thus {\it half} a regular 1D Fermi gas,
or a non chiral ``Majorana quantum wire".

Below it will be useful to consider
junctions that bend and close.  When a line junction makes an angle
$\theta$ with the $x$ axis the basis vectors (\ref{zeta}) are
modified according to $\zeta_a\rightarrow e^{i \sigma_z
\theta/2} \zeta_a$.  $\tilde{\cal H}$, however, is
unchanged even when $\theta(x)$ varies.  On a circle,
$\zeta_a$ changes sign when
$\theta$ advances by $2\pi$.  Therefore,
eigenstates of $\tilde{\cal H}$ must obey {\it antiperiodic} boundary
conditions, $\varphi(0) = -\varphi(2\pi)$.

Next consider a tri-junction, where three superconductors
separated by line junctions meet at a point, as in Fig. 1c.
When $\phi_{k=1,2}$ is in the shaded region of Fig. 1d, a MBS
exists at the junction.  Though the general BdG
equation cannot be solved analytically, this phase
diagram can be deduced by solving
special limits.  When $\phi_k = 0$ there is
no bound state.  Another solvable limit
is when three line junctions with $W=0$
are oriented at 120$^\circ$, and $\phi_k =\pm k (2\pi/3)$.  This is
a discrete analog of a $\pm$ vortex with $C_3$ symmetry, and is
indicated by the circles in Fig. 1d.  For $\mu = 0$ we find a MBS
identical to (\ref{vortex})
with the exponent replaced by $-\Delta_0 \hat n\cdot {\bf r}/v$.  Here
$\hat n$ is a constant unit vector in each superconductor
that bisects the angle between neighboring junctions.
The MBS can not disappear
when $\phi_k$ are changed continuously
unless the energy gap closes. The phase boundaries indicated in Fig.
1d therefore follow from the solution of the
line junction, and occur when the phase difference between
neighboring superconductors is $\pi$.

It is instructive to consider the limit where two of the lines
entering the tri-junction are nearly gapless.  For
$\phi_1=\pi-\epsilon_1$ and $\phi_2 =
\pi-\epsilon_2$ Fig. 1d predicts
a MBS when $\epsilon_1\epsilon_2<0$.
This can be understood with Eq. \ref{h1d}, which describes the lower two
line junctions, which have masses
$\delta_{1,2} = \Delta_0 \epsilon_{1,2}/2$.  When $\epsilon_1\epsilon_2<0$
$\delta$ changes sign, leading
to the well known midgap
state of the SSH model\cite{ssh,jackiw},
which in the present context is a MBS.

A line junction terminated by two tri-junctions
allows MBSs to be created, manipulated and fused.  When $\phi$
passes through $\pi$ MBSs appear or disappear at both ends.
To model this we assume the phases of the superconductors
on either side of the line junction are $0$ and
$\pi - \epsilon$, and that
the superconductors at the left (right) ends have phases
$\phi_{L(R)}$, which are not close to $0$ or $\pi$.  This allows us to
model the ends using a hard wall boundary condition
$\delta \rightarrow \pm \infty$, where the sign at each end is
$s_{L,R} = {\rm sgn}\sin\phi_{L,R}$.  It is straightforward to solve (\ref{h1d})
to determine the spectrum as a function of $\delta = \Delta_0 \epsilon/2$
for a line of length $L$ using this boundary condition.  There are two
cases depending on the sign of $s_L s_R$.

\begin{figure}
 \centerline{ \epsfig{figure=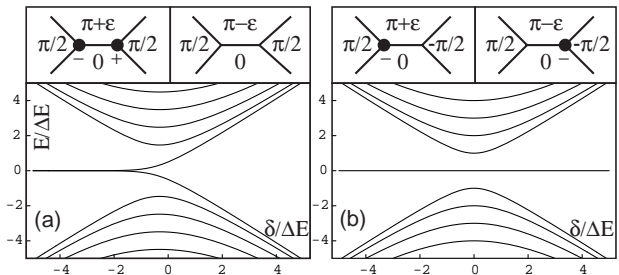,width=3.2in} }
 \caption{Energy levels in units of $\Delta E = \tilde v/L$ for a STIS line junction terminated by two tri-junctions as a
 function of $\delta = \Delta_0 \cos \phi/2$ for $\phi\sim \pi$.  In (a) two MBSs are created
 or fused when $\phi$ passes through $\pi$.  In (b) a single MBS is transported from one
 end to the other.  The insets depict the MBSs. }
\end{figure}

For
$s_L=s_R=1$ either zero or a $\pm$ pair of MBSs are expected.  The
spectrum, shown in Fig. 2a, may be written
$E_n^\pm(\delta) = \pm \sqrt{\delta^2 + \tilde v^2 k_n^2}$, where $k_n$ are
solutions to $\tan k_n L = - \tilde v k_n/\delta$.
Midgap states are present for $\delta  < 0$.  For
$L\rightarrow\infty$ a pair
of zero energy states $\xi_{1,2}(x,y)$
are localized at each end with wavefunctions
\begin{equation}
\xi_1 = \zeta_1  e^{-|\delta|
x/\tilde v},  \quad \xi_2  = \zeta_2  e^{-|\delta|
(L-x)/\tilde v},
\end{equation}
where $\zeta_a(y)$ are given in (\ref{zeta}).
For finite $-\delta L/\tilde v\gg 1$ the eigenstates are
$\varphi_\pm = \xi_1 \pm i \xi_2$, with energies
$E_0^\pm(\delta) \sim \pm 2|\delta| e^{-|\delta| L/\tilde v}$.
These define Bogoliubov quasiparticle operators,
$\Gamma_\pm = (\varphi_\pm)^\dagger \Psi$.  Since $\varphi_\pm = \Xi \varphi_\mp$
it follows that $\Gamma_+ = \Gamma_-^\dagger \equiv (\gamma_1-i\gamma_2)/2$
where $\gamma_a = (\xi_a)^\dagger \Psi$ are Majorana operators.
The pair $\gamma_{1,2}$ thus define a two state Hilbert space
indexed by $n_{12} = \Gamma_+^\dagger \Gamma_+$.
The splitting
between $\varphi_\pm$ then characterizes the interaction between
the MBSs,
\begin{equation}
H = E_0^+(\delta) ( \Gamma_+^\dagger\Gamma_+-1/2) = i E_0^+(\delta) \gamma_2 \gamma_1/2.
\label{splitting}
\end{equation}
The $s_L=s_R=-1$ case is similar.  Eq. \ref{splitting}
applies to both cases, provided $\gamma_2$ is associated with the $+$
vortex.

This provides a method for both creating and fusing pairs of
MBSs.  Suppose we begin in the ground state at $\delta >
0$ with no MBSs present.  Upon adiabatically decreasing
$\delta$ through $0$ MBSs appear in
the state $|0_{12}\rangle$.   Next suppose that initially $\delta<0$, and
 a pair of MBSs are present in the state $|n_{12}\rangle$.
When $\delta$ is
adiabatically increased through $0$ the system will remain in
$|n_{12}\rangle$, which will either evolve to
the ground state or to a state with one extra
fermion.  The difference between the two states can be probed by
measuring the {\it current} flowing across the linear junction,
which depends on whether the Andreev bound state $\varphi_+$ is occupied.
The measured current will be $I = \bar I \pm \Delta I/2$, where
the current carried by $\varphi_+$ is
$\Delta I = (e/\hbar)dE_0^+/d\phi \sim e \Delta_0/2\hbar$
for $\delta L/\tilde v \gtrsim 1$.  For $\Delta_0 \sim .1$ meV
$\Delta I \sim 10$ nA.

Finally, consider the case
$s_L = -s_R = 1$, in which a $-$ MBS is at one end or the other, as in
Fig. 2b.  There are plane wave solutions with energy
$E^\pm_n = \pm \sqrt{\delta^2 + (n \pi \tilde v/L)^2}$ for
$n=1, 2, ...$, along with a single $E_0=0$ state
with wavefunction
\begin{equation}
\xi(x,y) \propto
\zeta_1(y) e^{\delta x/\tilde v}
\end{equation}
Depending on the sign of $\delta$,
$\xi$ is exponentially localized at one end or the other.
When $\delta$ changes sign, the MBS
smoothly switches sides.  This provides a method for transporting
a MBS from one node to another.

We now discuss simple circuits built from
STIS junctions.  First, consider Fig. 3a and a
process in which the phase of the central island is
adiabatically advanced from $0$ to $2\pi$.  For $\phi=0$ there
are no MBSs.  At $\phi = 2\pi/3$ two pairs of
MBSs are created at the top and bottom line junctions.
At $\phi = 4\pi/3$ the
MBSs are fused at the left and right line junctions.
If the system begins in the ground state $\phi=0$, then when
$\phi\rightarrow 2\pi$ we find\cite{kitaev3}
\begin{equation}
|0_{12}0_{34}\rangle \rightarrow (|0_{14}0_{32}\rangle +
|1_{14}1_{32}\rangle)/\sqrt{2}.
\label{entangle}
\end{equation}
Thus, after the cycle, the left and right segments are in an
entangled state.  The currents measured across the left and
right junctions will be $\bar I \pm \Delta I/2$ with 50\% probability and will be
perfectly correlated.

\begin{figure}
 \centerline{ \epsfig{figure=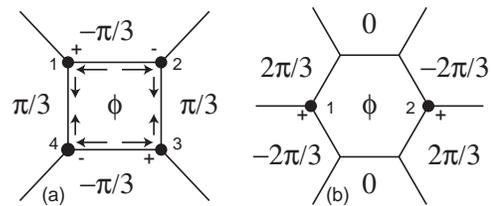,width=2.5
in} }
 \caption{Simple circuits made from STIS junctions.  When
 $\phi$ is advanced from $0$ to $2\pi$ (a)
 produces an entangled state and
 (b) interchanges MBSs 1 and 2.}
\end{figure}

Eq. \ref{entangle} can be understood in two ways.  First, the cycle
effectively creates two pairs of MBS's, interchanges a pair (say 2
and 4) and brings the pairs back together.  As shown by Ivanov
\cite{ivanov}, this corresponds to the operator
$P_{24} = (1+\gamma_2\gamma_4)/\sqrt{2}$, which leads directly to
(\ref{entangle}).  Alternatively, this result can be derived from
(\ref{h1d},\ref{splitting}).  From (\ref{splitting}),
the Hamiltonian for $\phi \lesssim 2\pi/3$ is
$H_1 \propto i(\gamma_1\gamma_2+\gamma_3\gamma_4)$.  For $\phi \gtrsim 4\pi/3$
it becomes $H_2 \propto i(\gamma_1\gamma_4 - \gamma_3\gamma_2)$.
Here the minus sign arises because, as explained after Eq.
\ref{h1d}, the closed 1D circuit must have antiperiodic boundary
conditions.  Thus, one of the line junctions must have a cut where
the wavefunction changes sign.  We chose the cut to be on the
junction between 2 and 3.  It is then straightforward to
express the groundstate of $H_1$ in terms of the eigenstates of
$H_2$, which leads directly to (\ref{entangle}).

Fig. 3b gives a geometry for interchanging MBSs without fusing them.
For $\phi = 0$ MBSs are located as shown.  When $\phi$
advances by $2\pi$ the MBSs hop counterclockwise three
times and are interchanged.  Ivanov's rules\cite{ivanov} predict
$\gamma_2 \rightarrow \gamma_1$, $\gamma_1 \rightarrow -\gamma_2$.
Again the minus sign can be understood in terms of the cut due to
antiperiodic boundary conditions.  One can imagine larger arrays,
where this process performs elementary braiding operations.

The experimental implementation of this proposal will require
progress on many fronts.  The first is to find a strong TI
with a robust gap.  Bi$_{1-x}$ Sb$_x$ and strained HgTe can
have gaps of order 30 meV \cite{fukane}. The next is to interface
with an appropriate superconductor. $\Delta_0$ depends on the quality
of the interface, Schottky barriers and the mismatch in the Fermi
wavelengths\cite{volkov}.  If these can be optimized, $\Delta_0$ can
be comparable to the gap of the bulk superconductor \cite{chrestin}.

The simplest experimental geometry
would be to consider a single line junction with $W \lesssim \hbar v/\Delta_0$.
For $\Delta_0\sim$ .1 meV and $\hbar v\sim 1$ eV\AA\ this could be achieved
with $W \lesssim 1 \mu$m.
This should
be similar to a graphene SNS junction\cite{beenakker}.
A signature of the Majorana character of the junction could be probed
by measuring the thermal conductance along the channel for $k_BT < \Delta_0$.
For $\phi=\pi$ the central charge $c=1/2$ of the
gapless Majorana modes leads to a quantized Landauer
thermal conductance $\kappa = c (\pi^2/3)(k_B^2/h)T$.
By constructing a pair of
tri-junctions as in Fig. 2 the presence of
MBSs can be controlled.  It would then
be interesting to perform tests of the non locality of
MBSs envisioned in Refs. \onlinecite{semenoff,demler}.

Manipulating and fusing MBSs places more
stringent requirements on the energy gaps.  The junctions should
be sufficiently short that
$\Delta E = \tilde v/L > k_B T$, but sufficiently long
that the MBSs are well localized.  The good news is that
$\Delta E$ varies as a power of $L$, while the MBS overlap
is exponential, so at low temperature both criteria can
be achieved.

If the process of varying $\delta$ to manipulate the MBSs
is non adiabatic or $\Delta E < k_B T$ then
additional quasiparticles could be excited.
If those quasiparticles escape and interact with
other MBSs, then the state of the MBSs will
be disturbed.  However, if $\delta \ll \Delta_0$ the excited
quasiparticles will be confined to the segment in which they were
created.  If $\delta$ is turned up so that
$k_B T \ll \delta \ll \Delta_0$, and
the system relaxes back to its ground state, then the state of
the MBSs will remain intact.  Thus, if there is
sufficient dynamic range between $k_B T$ and $\Delta_0$, the system
can tolerate these excitations.


We thank Sankar das Sarma and Steve Simon for
helpful discussions.   This work was supported by NSF grant
DMR-0605066, and by ACS PRF grant 44776-AC10.

\end{document}